\documentstyle[aas2pp4]{article}
\begin{document}
\newcommand{\etal}{{\it et~al.\ }}
\title{Spectroscopic properties, spatial and luminosity distributions\\ 
of the UCM survey galaxies\altaffilmark{1}}

\author{J. Gallego\altaffilmark{2}}
\affil{Lick Observatory, University of California, Santa Cruz, CA
95064, USA; jgm@ucolick.org}
\affil{Dpto Astrof\'{\i}sica, Universidad Complutense de Madrid,
    E-28040 Madrid, Spain}
\author{J. Zamorano, M. Rego}
\affil{Dpto Astrof\'{\i}sica, Universidad Complutense de Madrid,
    E-28040 Madrid, Spain; jaz@ucmast.fis.ucm.es}
\and
\author{A.G. Vitores}
\affil{Dpto Astrof\'{\i}sica, Universidad Complutense de Madrid,
    E-28040 Madrid, Spain}
\affil{EUIT Industrial. Universidad Polit\'{e}cnica, E-28012, 
Madrid, Spain}

\altaffiltext{1}{Partly based on observations collected at the German-Spanish 
Astronomical Center, Calar Alto, Spain, operated by the Max-Planck-Institute 
f\"{u}r Astronomie (MPIA), Heidelberg, jointly with the Spanish National 
Commission for Astronomy. \\
Partly based on observations made with the Isaac Newton Telescope operated on 
the island of La Palma by the Royal Greenwich Observatory in the Spanish 
Observatorio del Roque de Los Muchachos of the Instituto de Astrof\'{\i}sica 
de Canarias.}
\altaffiltext{2}{del Amo UCM Foundation Fellow.}
\begin{abstract}
A spectroscopic analysis of the whole sample of H$\alpha$
Emission-Line Galaxies (ELGs) contained in the lists 1 \& 2 of the
Universidad Complutense de Madrid (UCM) objective-prism survey is
presented. A significant fraction (59\%) of star-forming galaxies with
low ionization or high extinction properties has been found.  This
kind of ELG is only incompletely detected in the blue or in other ELG
surveys.  We have found evidence for evolution among some of the
different ELG classes.  A comparison between the populations
detected by Case, Kiso, UM and UCM surveys is presented. We conclude
that a deep H$\alpha$ survey is better able to sample all the ages,
evolutionary stages and luminosities of current star-forming galaxies
than other surveys using blue-emission lines or colors.  Finally, the
luminosity and spatial distributions of the UCM galaxies are
determined. The contribution of the newly found current star-forming
ELGs adds new clues to galaxy evolution and has to be taken into
account when trying to consider the density of ELGs 
and total Star Formation Rate in the Universe.
\end{abstract}


\section{Introduction}
The UCM objective-prism survey is being carried out with the main
purposes of identifying and studying new star-forming galaxies and
quantifying the properties of the Star Formation Rate (SFR) in the
local universe. The technique used for detection is the presence of
H$\alpha$+[NII] $\lambda$6584 emission at objective-prism Schmidt
photographic plates. Because the specific details have already been
summarized in the first and second lists
(Zamorano \etal 1994;1996) of our exploration, and the photometric aspects
of the sample are presented in Vitores \etal (1996a and 1996b), we
concentrate here in the spectroscopic characteristics of the survey.
A companion study of the far-infrared properties of this sample of
ELGs is published in Rego \etal (1993). Finally, the H$\alpha$
luminosity function and the current SFR from a
complete sample of the survey can be found in Gallego \etal (1995).

We have carried out a program for obtaining long-slit CCD spectroscopy
of the whole lists 1 \& 2 of the UCM survey.  The main goals are (1)
the selection characteristics, biases, and completeness limits of the
sample, (2) the physical properties that are responsible for the
spectroscopic behavior and (3) the spatial and luminosity
distributions to determine how they compare with those of other ELGs
surveys and normal galaxies.  The spectroscopic data and individual
classifications are available in Gallego \etal (1996), hereafter
referred as Paper I.

Given the importance of the comparison between the results of such a
survey in the red and others using different techniques for detecting
ELGs we have also considered those surveys with a similar analysis
available in the literature.  As a representative sample of blue
objective-prism surveys we have used the lists III and IV of the
University of Michigan (UM) survey (Salzer 1989, Salzer \etal 1989a \&
Salzer \etal 1989b). For the surveys based in the blue excess we have
used the subsample of Kiso Ultraviolet Galaxies (KUG) from Augarde
\etal (1994) and Comte \etal (1994).  Finally, as a mix of blue
objective prism and blue excess we have used the Case survey for blue
or emission-line galaxies (CG) from Rosenberg \etal (1994) and Salzer
\etal (1995).

The current paper presents a discussion on the overall properties of
the UCM galaxies.  In section~2 we address the physical properties.
In section~3 we focus on spatial distributions.  The selection effects
are considered in section~4. Finally, we present a discussion about
the luminosity function in section~5. A summary of our results and
conclusions is given in section~6. Unless it is specified a value 
of H$_{\rm o}$=50 km s$^{-1}$ Mpc$^{-1}$ was assumed. 

\section{Physical properties of the UCM emission-line galaxies}

In order to classify the ELGs found by the UCM survey, we have adopted
the scheme of Salzer \etal (1989a, 1989b).  The reader is referred to
paper I for a complete description of the various types. We have only
considered the observationally well-defined Seyfert 1 (Sy 1) and
Seyfert 2 (Sy 2) galaxies. Also were considered Starburst Nuclei
(SBN). These are spiral galaxies that host a nucleus with an
star-forming process. The H$\alpha$ luminosity is greater than 10$^8$
L$_{\odot}$ and present red colors.  Any emission line present in the
blue are faint due to extinction.  The Dwarf Amorphous Nuclear
Starburst (DANS) are similar to the SBN but at lower scale. They are
spectroscopically indistinguishable from SBNs, but their H$\alpha$
luminosities are lower than 5 10$^7$ L$_{\odot}$.  The HII Hotspot
class (HIIH) includes all those bright galaxies with a global
star-forming process and an optical spectrum dominated by blue colors
and strong emission lines.  Their H$\alpha$ luminosities are similar
to those of SBN class.  The Dwarf HII Hotspot (DHIIH) galaxies have
spectroscopic properties similar to those of HIIH class except
H$\alpha$ luminosities lower than 5 10$^7$ L$_{\odot}$. Finally, the
Blue Com\-pact Dwarfs (BCD) are characterized by H$\alpha$
luminosities $\leq$ 5 10$^7$ L$_{\odot}$, strong emission lines and
equivalent widths (EW) of several hundred angstroms. Giant Irregular,
Magellanic Irregular and Interacting Pair classes were not considered
because these types rely mainly on morphological aspects of the
candidate.

\subsection{Overall spectroscopic properties} 
Some spectroscopic and physical properties for each of the UCM ELGs
types are presented in tables \ref{tab-cocmed} and
\ref{tab-parmed}. Table \ref{tab-cocmed} lists the mean values and
standard deviations of the [OIII] $\lambda$5007/H$\beta$, [NII]
$\lambda$6584/H$\alpha$, [OII] $\lambda$3727/H$\beta$, and [OI]
$\lambda$6300/H$\alpha$ emission-line ratios. The last column gives
the number of ELGs.  Table \ref{tab-parmed} lists the mean values of
several fundamental parameters.
The composition of the sample by types is 14 Seyfert galaxies (8\%),
111 red ELGs (SBN and DANS, 59\%) and 61 blue HII-spectrum ELGs
(i.e. HIIH, DHIIH and BCD, 33\%).  It is worth noting that meanwhile
the UM survey contains 11\% and the Case survey 18\% of SBN galaxies,
the UCM survey is detecting 44\%. Since SBN have weak lines in the
blue, all this population of star-forming galaxies is being poorly
considered by surveys in that spectral region.  Lewis (1981) called
them WOR (Weak Oxygen Red) galaxies. Salzer \etal (1989) includes only five
objects in the SBN class for the UM survey.  However, almost all of
the H$\alpha$ luminosity observed in the UCM survey at bright
absolute-magnitudes is hosted in the form of SBNs (Gallego \etal 1995,
Coziol 1996). Salzer \etal (1989a) claim that the UM survey would
recover all the SBNs by the [OII] $\lambda$3727 line, but 38 SBN of
the UCM survey do not present any [OII] in emission and the mean EW is
only 12 \AA \ (27 \AA \ when considering only the 45 with [OII]
detected). Because the contrast limits for both UM and UCM surveys are
10 \AA, the completeness when using this line can not be assured for
the SBN class. The CFRS survey (Tresse \etal 1996) has also found a
considerable 38\% of SBN galaxies in a field sample up to z=0.3.

The DANS galaxies are also poorly sampled by the UM survey (a total of
15 objects in list III and IV, 0.04 $\sq^{-1}$) but are well
represented in both Case (37 objects, 0.20 $\sq^{-1}$) and UCM (28
objects, 0.08 $\sq^{-1}$) samples.  They present almost identical
characteristics no matter the survey considered, with weak lines in
the blue, average M$_B$=-18.4 and EW([OIII])=$\lesssim$10 \AA.  For
the UCM sample we obtain an average EW(H$\alpha$)=54 \AA \ and average
colors of $\bv$=0.38 and $\vr$=0.28 with E$_{\bv}$=0.55. These
properties point to reddened low-ionization objects easily found by
the H$\alpha$+[NII] emission but only detectable by the [OII]
$\lambda$3727 line (38 \AA \ mean value for the UCM) at the
blue-excess or blue emission-line surveys. The mean H$\alpha$
luminosity obtained for the DANS UCM subsample is only 0.58 10$^{8}$
L$_{\odot}$ but, as it will be seen in the next section, present ELGs
surveys could be missing a large fraction of such objects.

It is worth noting that the BCDs in the UCM survey are in average an
absolute magnitude brighter than those found by UM and Case
surveys. BCDs are characterized by faint absolute magnitudes and large
equivalent widths, with the largest values for the lowest absolute
magnitudes.  The apparent magnitude limit of the UCM survey is close
to r=18 (Vitores \etal 1996b), instead of the 19.5 apparent magnitude
for the UM survey. Even at very low redshift, galaxies with absolute
magnitudes below -16 would appear fainter than the UCM limit value
(for z=0.01 and M$_r$=-15.5 it results an apparent magnitude of
r=18.5).  In fact, the UCM survey presents a lack of sensitivity for
low luminosity objects, detecting only the brightest members of the
BCD class, i.e., those with EW less extreme (The faintest object is
UCM1612+1309, with M$_r$=-16.7).  Furthermore, the average EW([OIII])
for the UM, UCM and Case BCD subsamples are 1090, 458 and 247 \AA,
confirming that the UCM value is biassed by its lower sensitivity to
faint absolute magnitude and large EW objects.

\subsection{Line-ratio and ionization diagrams} 
In Figure \ref{f1-diag} has been represented the [NII]/H$\alpha$
versus [OIII]/H$\beta$ diagnostic diagram (Veilleux \& Osterbrock
1987) for the UCM sample. As expected, all the ELGs classes are
positioned following a narrow sequence defined by the HII models.  It
starts at the upper-left corner of the diagram for the high-ionization
low-metallicity BCDs, and it ends at the lower-right corner for the
low-ionization and high-metallicity objects.

\placefigure{f1-diag}

The low ionization corner is well populated by the UCM galaxies. In
comparison with the UM survey, the UCM survey is recovering a higher
fraction of low ionization galaxies.  The comparison is not
straightforward for the Kiso and Case samples because only a
relatively small number of galaxies from both surveys were observed
spectroscopically in the H$\alpha$ region.

The low-ionization population is also present in the Figure
\ref{f2-o3mb}. It plots the logarithm of the [OIII]/H$\beta$ ratio
(the excitation parameter) against absolute magnitude M$_B$. As a
complement for the UCM sample we have added the
eleven objects found by Boroson \etal (1993, hereafter referred as
BST) in a search for extremely low-luminosity objects using H$\alpha$
and narrow-band filters at the KPNO 0.9m telescope.  This completes
the low-luminosity end of the H$\alpha$-selected sample.

\placefigure{f2-o3mb}

ELG's spectrum is the result of the physical
properties and the importance of the on-going starburst relative to
the underlying population.  Based in time scale arguments, Salzer et
al. (1995) pointed out that there should be a large population of
low-ionization dwarf galaxies which starburst is already after the
peak in luminosity and that is not well detected by the UM or any
other existing survey.  These objects would show a soft spectrum with
weak emission lines and small EWs.  According to the
models (Leitherer \& Heckman 1995, Stasinska \& Leitherer 1996), if we
let a starburst evolve several Myr, they move along the HII sequence
to the lower-right corner in Figure
\ref{f1-diag}, but to the lower-left in Figure \ref{f2-o3mb}, out of the HII
sequence. The properties of the pure burst remain almost constant
until the 10Myr step, when the O stars begin to disappear, a drop
happens in all physical properties and the object presents a weaker
emission-line spectrum.  M$_{B}$ becomes 1-2 magnitudes fainter,
$\Delta(\vr)\sim$0.4 mag and EW(H$\alpha$) drops by a factor of ten.
Detection techniques based in blue colors or H$\alpha$ emission (as
Case or UCM surveys), have advantage over those using
high-ionization lines (as UM survey) when detecting these
soft-spectrum galaxies.  A fraction of such a population is actually
being recovered in the form of DANS objects.  At Figure \ref{f2-o3mb}
these objects have been oversized for better recognition. They are
low-ionization ELGs, with low EWs and a lower M$_B$ than
the expected value from the HII sequence. All these magnitudes are
coherent with a past-the-peak stage.  Furthermore, the parental
population for DANS galaxies would be ELGs with higher ionization
features and larger EWs, when the ionizing stars have a higher mean
effective temperature. This pre-DANS population can be found among the
HIIH and DHIIH ELGs, whose physical properties are the expected.

\placefigure{f3-exci}

 In Figure \ref{f3-exci} the ionization diagram (log([OIII]/H$\beta$)
versus log ([OII]/[OIII])) for the UM, KUG, UCM and Case galaxies is
shown. The models from Stasinska \& Leitherer (1996) are plotted for
different metallicities and the whole range of ionization
parameters. High ionization is in the upper-left and low ionization is
in the lower-right.  As expected, only some of the possibilities are
populated by ELGs.  At the high ionization region all the samples
except the KUG one are well represented.  Only Case and UCM present
galaxies than deviate from the HII sequence, being mainly in the
low-ionization region.

For explaining the better ability of the UCM survey to recover all
kind of star-forming galaxies we have to rely on the nature of the
tracer used by each survey.  Since it is directly related to the
number of massive stars, the H$\alpha$ luminosity is a direct
measurement of current SFR.  It is better than other optical Balmer
lines like H$\beta$ ---affected by stellar absorption and reddening and
with smaller photon flux---.  Metallic nebular lines like [OII],
[OIII] ---affected by excitation and metallicity--- IRAS fluxes
(affected by the dust abundance and properties) or broad-band
luminosity densities ---dependent of stellar libraries for
calibration and very sensitive to the underlying stellar
population--- are more star-formation indicators than quantitative
measurements (see, eg., Gallagher \etal  1989 and Kennicutt, 1992).
These considerations imply that the best way to trace and quantify
current star formation processes in the whole range of physical
properties is by using an H$\alpha$-based detection technique. As a
H$\alpha$-selected sample, the UCM sample populates almost all the
regions of the ionization diagram.

\subsection{Emission-line equivalent widths}
We present in Table \ref{tab-ewoiii} and in Figure \ref{f4-ewoiii} the
distribution in [OIII] EWs for the UCM sample, the Kiso data from
Comte \etal (1994), and the UM data from Salzer \etal (1989a).  The
same data for H$\beta$ EWs can be found in Table \ref{tab-ewhb}.
 
\placetable{tab-ewoiii}
\placetable{tab-ewhb} 

A total of 44\% of UCMs do not present [OIII] at all, whereas 19\%
present EWs greater than 100 \AA \ (age of a pure starburst below 5
Myr).  The KUG sample has also a high fraction of no [OIII] galaxies
(34\%) but only has a 5\% of objects above the 100 \AA \ limit.

\placefigure{f4-ewoiii}

The UM sample is so rich in objects with large EWs 
that the mean value of the is almost 100 \AA.  
The fraction of ELGs with no line is very
low (4\%), because this survey detects the candidates mainly by the
presence of this line. Again the UM sample seems to be more biassed to
high ionization objects, missing a significative fraction of
low-ionization star-forming galaxies. The reason why the Kiso sample
does not detect large numbers of high-EW objects remains unclear. It
may be due to a bright value in the apparent magnitude limit or to a
selection effect in the subsample considered by Comte \etal (1994).

\placefigure{f5-ewsurveys}

Because of its mixed blue excess and blue emission-lines selection
nature, the Case survey should detect the full range of EWs.  
Now in Figure \ref{f5-ewsurveys} the [OIII] EW histograms are
plotted in a different way. A rectangle delimits the position of the
first and third quartile and the mean value is also marked. Finally
the whole range covered is delimited by a line. The mean and third
quartile for Case and UCM samples agree pretty well, meanwhile the UM
sample is centered at higher values. 

Finally, one of the most fundamental parameters for characterizing the
population of UCM ELGs is the distribution
according to the EW(H$\alpha$+[NII]) (Figure \ref{f6-ewha}).

\placefigure{f6-ewha}

The mean value reaches 102 \AA, with 35\% of the objects above this
value. In the high-EW end there are three objects (UCM0056+0044,
UCM1331+2901 and UCM1612+1309) with values over 400 \AA. In the low-EW
end only 15 objects (8\%) are below 20 \AA. Meanwhile no values over
1000 \AA \ are set by the roles of the hottest plausible O-star
continua and significant nebular continuum, it is clear that below the
10 \AA \ limit the sensitivity of the survey decreases steeply. The
wide range covered in the distribution points out that the H$\alpha$
technique is able to detect star-forming galaxies in an universal way.

\subsection{Abundances of the UCM galaxies}
In order to estimate the metallicities of the UCM galaxies, we 
contrasted the emission-line ratios available for 125 UCM galaxies
with the Stasinska \& Leitherer (1996) stellar evolutionary synthesis
code and single-zone gas (spherical symmetry, uniform chemical
composition and density distribution) nebular photo-ionization code
(Stasinska 1990).  The results are summarized
in Table \ref{tab-abund}.  For each class it is given the number of
components within the different metallicities considered, the total
number and the mean metallicity in logarithmic (the solar abundance 
corresponds to 8.82) and natural solar units. 

\placetable{tab-abund}

The redder SBN and DANS
classes are more metallic than the bluer HIIH, DHIIH and
BCDs (as expected from the self-enrichment by the
current starburst episode if they correspond to a more
evolved stage). We should find extremely metal-poor
galaxies between the subsample of objects classified as BCDs. However,
we also would expect to observe low metallicities at the
low-ionization dwarfs discussed before. Perhaps the presence of two
DANS with 1/10 Z$_{\odot}$ points to this possibility but more
metallicities of objects similar to a4.483 are needed.

\placetable{tab-candida}

We also carried out high signal-to-noise spectrophotometry for a
subsample of 15 UCM galaxies.  The procedure followed is step-by-step
described by Pagel \etal  (1992). The average error for the
electronic temperature is 5\%.  Considering the problem of the
primordial Helium, when the electronic temperature was available and
the He line fluxes were accurate enough (within a 10\% error), also
the Helium abundance was estimated.  In Table \ref{tab-candida} we
give the electronic density as computed from the [SII] lines,
the electronic temperature, the Oxygen abundance, the Oxygen to
Nitrogen ratio, the ionic O$^{++}$/Ne$^{++}$ and S$^{+}$/H$^{+}$
ratios and finally the Helium abundance.  Using these values, the
metallicities estimated from the models are correct within a 25\%.
None of the galaxies found have lower metallicities than 1/20
Z$_{\odot}$. The reasons are mainly two. The first one is that it
could be expected that, if galaxies with very low metallicities exist,
they have very low luminosities, so only a ultra-deep survey would
find them. The second one is a self-contamination problem (Kunth \&
Sargent 1986).  There must be young massive O and B stars for having
the emission-line spectrum used for computing the abundances. But
these stars have strong winds that inject higher metallicity material
in the interstellar medium. If so, the lowest abundances would be
found at the neutral gas out of the already contaminated HII region
surrounding the ionizing stars. This hypothesis seems to be confirmed
by Kunth \etal (1994).

\section{Spatial distribution of the UCM galaxies}

\subsection{Pie diagrams and clustering}

We utilized the CfA data (Huchra, Geller, Corwin 1995 and references
therein) for  the redshifts of all galaxies
previously known at the regions covered by the UCM survey. A total of
4219 CfA galaxies and 196 UCM ELGs have been plotted in Figure
\ref{f7-tartas}.  The declination dimension was suppressed.

\placefigure{f7-tartas}

The top panel corresponds to a highly clustered region centered at
$\alpha$=15$^{\rm h}$ (UCM list 2).  It includes the Coma
(R.A.$\sim$13$^{\rm h}$) and the Hercules clusters (R.A.$\sim$16$^{\rm
h}$).  The bottom panel corresponds to a region of moderate density of
field galaxies centered at $\alpha$=0$^{\rm h}$ (UCM list 1).  Whereas
in the Coma cluster there is a large number of H$\alpha$ emission-line
galaxies it is worth noting that none of the Hercules galaxies were
detected.  We leave here this question open but perhaps it would be
interesting a deeper study in H$\alpha$ of this cluster.

\placefigure{f8-cfa228}

In Figure \ref{f8-cfa228} a projection on the sky with all the
galaxies that belong to the Coma cluster is shown.  A total of 18 UCM
ELGs inside Coma were not previously known, suggesting that perhaps
our actual knowledge of the Coma population is not so complete.  In
fact three of each four are DANS or SBN and consequently reddened
galaxies. How many of these galaxies remain undiscovered at Coma is a
question being only slowly resolved (see for example Caldwell et
al. 1996).  The two-point correlation function for both all cluster
(above, dots) and UCM samples (below, open symbols) are also
plotted. The UCM galaxies (i.e., the star-forming galaxies) at the
Coma cluster are less clustered than the galaxies of the cluster as a
whole.  This picture suggests the idea that the star-formation inside
clusters is indirectly modulated by the density of the intergalactic
medium.

\section{Completeness of the UCM survey}
In any survey for detecting emission-line features in objective-prim
plates the apparent magnitude is not the only factor that contributes
to the selection of a candidate.
\subsection{Total flux of the line+continuum feature.} 
Any candidate needs a total flux in
the H$\alpha$+[NII] region over a threshold value in
order to be registered by the plate emulsion. The total flux and the
equivalent width are two  fundamental parameters for
defining the selection criteria.

\placefigure{f9-ewhar} 

In the Figure \ref{f9-ewhar} both parameters are plotted for the UCM
sample. It can be observed a trend of larger EWs for fainter ELGs.
According to the continuum decreases the line flux has to
increase in order to maintain the line+continuum over
the threshold value. The fundamental parameter of detectability is
this total flux. In case of an universal population of galaxies all
the regions in Figure \ref{f9-ewhar} might be occupied.  In our Figure
there are empty regions. Is very easy that a near-saturation continuum
masks any possible emission (lower-left corner). As an example the
object NGC7677 is a 13.9$^{\rm m}$ nearby galaxy with well-known
emission-lines.  However, this galaxy presents a completely saturated
spectrum with no detectable emission. At lower-right corner are the
faint galaxies with small EWs.  Also the Surface
brightness becomes important when the spectrum is near the saturation
limit.  If the total luminosity of the candidate is spread across a
larger area it will be easier to notify the possible emission present.
Conversely, low-surface brightness galaxies will be lost if no bright
emitting knot is present.

\subsection{Contrast of the emission line over the continuum.}
Stronger emission lines make detection easier.  There is a threshold
value of $\sim$10 \AA \ below which no objects are detected (see
Figure \ref{f10-ewhaz}). Only two galaxies were selected with lower EW
values. The first was a miss-selection of an edge-on galaxy with a
superimposed star (UCM2320+2428, 7 \AA) and the second is a diffuse
low surface-brightness galaxy (UCM2249+2149, 4 \AA). An (x) marks the
position this object and of a8.196 
(not detected because of its faintness although the
EW is 84 \AA). Several galaxies with confirmed emission, 
but with no emission feature in the
objective-prism plates are labeled by crosses (+).  

\placefigure{f10-ewhaz} 

% \subsection{Redshift.} 
Because of the wavelength cut-off of the photographic emulsion, there
is an upper limit in redshift at 0.045$\pm$0.005. Since this cut-off
is not completely sharp, objects with strong emission are detected at
little higher redshifts (threshold $\sim$30 \AA \ at z=0.04).  The
variable dispersion of the prism, very important in the blue, 
 is negligible in the H$\alpha$ region.

\section{Luminosity function of the UCM galaxies} 
For comparison
purposes we have computed the luminosity function in a similar way
that Salzer (1989) did for the UM survey.  This author applies the
original V/Vmax method (Schmidt 1968, Huchra \& Sargent 1973) but
considering a synthetic magnitude for the total flux of line+continuum
($F_{{\rm L}+{\rm C}}$ in erg$\,$s$^{-1}\,$cm$^{-2}$) at the Schmidt
plate instead the normal apparent magnitude. This arbitrary magnitude
 is a function given by: 
\[ m_{{\rm L}+{\rm C}} =
-17.0 - 2.5 \log \: (F_{{\rm L}+{\rm C}}) \] 
where the total $F_{{\rm
L}+{\rm C}}$ line+continuum apparent flux comes from 
\[ F_{{\rm
L}+{\rm C}} = F_{\rm L} \left( 1 + \frac{2 P R}{EW} \right) \] 
where $
F_{\rm L}$ is the line flux and EW is the equivalent width (both
measured in the spectra), P is the reciprocal dispersion of the
objective prism
 and R is the spectral resolution (1950 \AA
\ mm$^{-1}$ and $\sim$ 10$\mu$m respectively). 
The quantity V/Vmax results
\[
\frac{V}{V_{max}} = {\left( \frac{r}{r^{*}} \right)}^{3} = 
10^{0.6 \: (m-m^{*})}
\] 
The subsample of 176 UCM galaxies with m$_{{\rm L}+{\rm C}}$ brighter
than 17.3 that makes \(\left< V/Vmax \right> = 1/2\)  and
EW(H$\alpha$+[NII]) $>$ 10\AA \ was considered
as complete and representative for the UCM galaxies. 
The $m_{{\rm L}+{\rm C}}\leq 17.3$ limit corresponds to a line+continuum 
flux of $1.9 \times 10^{-14}\,$erg$\,$s$^{-1}\,$cm$^{-2}$. Below this
value the sample presents larger incompleteness. 

Now the classical approach (Huchra \& Sargent 1973) was followed for
this sample. 
First, the value of Vmax is obtained from
\[ V_{max} = \frac{4}{3} \; \pi \; 10^{0.6 \: (m^{*} - M -25.0)} \]
where $m^*$=17.3 and M is the absolute magnitude
corresponding to the synthetic magnitude and computed
with the known redshifts and H$_o$=75 km s$^{-1}$ Mpc$^{-1}$.
The Schmidt estimator is then given by 
\[ \phi(M) =  \frac{4\pi}{\Omega} \sum_{i} \left( \frac{1}{V^{i}_{max}} 
\right) \] 
where the summation is over all galaxies with blue absolute
magnitude in the interval M$_B\pm$0.5 and $\Omega$ is the solid angle
in steradians covered by the survey.  We choose H$_o$=75 and the blue
absolute magnitude in order to compare our results with the other
surveys previously analyzed. The final luminosity function for the
UCM and the published values for BST, UM, Kiso and Case 
(UV-excess-selected only) galaxies are listed in Table \ref{tab-lf}.

\placetable{tab-lf}

The table gives log $\phi$(M) (galaxies per unit magnitude interval
per Mpc$^{3}$) and the number of galaxies included in each magnitude
bin. The same luminosity functions are displayed in Figure
\ref{f11-lf}. The errors bars plotted represent the square root of the
number of galaxies in each absolute magnitude. Symbols with no error
bars associated correspond to intervals which contain only one galaxy.
In the upper-left corner have been plotted BST (open diamonds) and UCM
(filled circles) luminosity functions together for a global
H$\alpha$-selected sample.  \placefigure{f11-lf}
 
We adopted a classical Schechter function (Schechter 1976) as given by
Felten (1977) where the free parameters are $\phi^{*}$, $M^{*}$ and
$\alpha$. The $\alpha$ corresponds to the slope of low-luminosity end,
whereas $\phi^{*}$ and $M^{*}$ give the position of the turning
point. The best fit to the data is given in Table \ref{tab-fsch}.
 
\placetable{tab-fsch}

The luminosity functions reflect what we have already pointed out in
previous sections. The low-luminosity end is poorly sampled by the UCM
survey due to the bright apparent magnitude limit. In any case, the
study of Boroson \etal (1993) becomes the perfect complement for the
UCM survey data, coinciding at the common bin of -18.  Joining the two
samples, the resulting luminosity function for H$\alpha$ emission-line
galaxies is the more accurate measurement of the luminosity
distribution of current star-forming galaxies.  This situation for
H$\alpha$ versus other lines is analogous to that in galactic nuclei,
where both AGN and star formation are more often detectable at
H$\alpha$ than in the blue (see for example Heckman 1980 or Keel
1983). It is worth noting that $\phi^{*}$ for the total H$\alpha$ ELGs
sample is larger than the Case value and twice the UM value. The total
density of ELGs rises from $\sum \log \phi (M)$=-1.51 for the UM ELGs
to $\sum \log \phi(M)$=-1.15.  When comparing this value with the
obtained from field luminosity functions, the H$\alpha$ ELGs
would amount for 0.07 galaxies per cubic Mpc, i.e., for $\sim$15\% of
all galaxies over the luminosity range considered.

\section{Summary and conclusions} 
We present the spectroscopic properties, spatial distribution and
luminosity function for the sample of H$\alpha$ ELGs from the lists 1
\& 2 of the UCM survey.  We have found a large fraction (59\%) of
low-ionization or high-extinction star-forming galaxies.  These
objects are poorly sampled by other surveys in the blue since they do
not present blue colors or strong emission lines.  However, as
these galaxies are current star-formers, they have to be taken into
account when considering the SFR in any volume of Universe surveyed.

We have found evidences for evolution between the different ELGs
classes considered. The DANS objects seem to be the past-the peak
stage for the HIIH and DHIIH classes. Several arguments including
total luminosity, equivalent widths and colors favor this hypothesis.
The extension of these objects towards low luminosities predicts a
population of low ionization and luminosity ELGs very difficult to
trace. The population of extremely-faint H$\alpha$ emission-line
galaxies found by BST can be considered as a probe of such population.
If even the UCM, UM or Case surveys are tracing these galaxies not
properly is because there are strong selection effects. In fact, the
Case survey detects several low-ionization candidates using the blue
color-excess, the UM detects those with higher ionization and the UCM
a mixing of both types.  We have estimated average abundances for
every ELG class. Accurate spectrophotometric abundances for several
objects confirm (within 25\% error) these results.  No galaxy with
metallicity below 1/20 Z$_{\odot}$ have been found.

Because the UCM galaxies are selected by their H$\alpha$ emission and
this feature is the best tracer for current SFR, we tried to compare
the distribution of the UCM and CfA samples 
in order to obtain new clues about the effect of the intergalactic
environment in the triggering of the starburst processes.  The sample
 covers both regions with low and high density of galaxies.  In the
Coma field, H$\alpha$ ELGs are considerably less clustered than the
normal population. This result points to a dependence of the SFR with
the galaxy density. A total amount of 18 not previously known
constituents were discovered in this well-studied cluster by means of
their H$\alpha$ emission. Also quite interesting is that no one H$\alpha$
ELG has been detected in the Hercules cluster.
 
The parameters that determine the selection of an object by the UCM
survey are the total line+continuum flux and the equivalent width of
H$\alpha$+[NII]. The UCM sample seems to be poorly sampling
low-luminosity galaxies.  The poor information for the low end of the
luminosity function was completed after considering the analysis
optimized for these objects by Boroson \etal (1993). As a global
result, the H$\alpha$ selection method is better able to detect
galaxies at any level of star-forming activity than previous
surveys.  The total density of ELGs rise to 0.07 galaxies per cubic
Mpc, i.e., roughly the 15\% of all galaxies over the luminosity range
considered and almost twice the value obtained by Salzer (1989) for
the UM survey.

\acknowledgments

This work was supported in part by the Spanish ''Pro\-grama Sec\-to\-rial de 
Promoci\'{o}n del Conocimiento'' under grants PB89--124 and PB93--456.
JG acknowledges the partial financial support from NASA grant GO-05994.01-94A.
He is grateful to D. C. Koo and R. Guzm\'an for their hospitality at 
Lick Observatory, UCSC.

\begin{deluxetable}{lccccccccr}
\tablewidth{0pc}
\tablecaption{Average line ratios for the UCM galaxies}
\tablehead{
\colhead{Type} & \colhead{[OIII]/H$\beta$} & \colhead{$\sigma$} 
& \colhead{[NII]/H$\alpha$} & \colhead{$\sigma$} & 
\colhead{[OII]/H$\beta$} & \colhead{$\sigma$} &
\colhead{[OI]/H$\alpha$} & \colhead{$\sigma$} & \colhead{N}
}
\startdata
Sy 1  & \makebox[0.3cm][r]{1.1} & 0.3 & \makebox[0.3cm][r]{0.3}  & 
\makebox[0.3cm][r]{0.2}  & \makebox[0.3cm][r]{1.7}  & 1.8 & 0.01 & 0.00 &  5\nl
Sy 2  & \makebox[0.3cm][r]{11.4}& 5.3 & \makebox[0.3cm][r]{1.5}  & 
\makebox[0.3cm][r]{1.1}  & \makebox[0.3cm][r]{11.0} & 7.9 & 0.20 & 0.20 &  9\nl
SBN   & \makebox[0.3cm][r]{1.5} & 1.1 & \makebox[0.3cm][r]{0.5}  & 
\makebox[0.3cm][r]{0.2}  & \makebox[0.3cm][r]{4.8}  & 3.5 & 0.10 & 0.10 & 83\nl
DANS  & \makebox[0.3cm][r]{1.6} & 1.1 & \makebox[0.3cm][r]{0.4}  & 
\makebox[0.3cm][r]{0.1}  & \makebox[0.3cm][r]{4.4}  & 2.9 & 0.03 & 0.01 & 28\nl
HIIH  & \makebox[0.3cm][r]{3.3} & 1.3 & \makebox[0.3cm][r]{0.2}  & 
\makebox[0.3cm][r]{0.1}  & \makebox[0.3cm][r]{4.5}  & 3.5 & 0.03 & 0.02 & 40\nl
DHIIH & \makebox[0.3cm][r]{4.0} & 1.0 & \makebox[0.3cm][r]{0.2}  & 
\makebox[0.3cm][r]{0.05} & \makebox[0.3cm][r]{5.3}  & 2.8 & 0.04 & 0.02 & 14\nl
BCD   & \makebox[0.3cm][r]{6.6} & 1.0 & \makebox[0.3cm][r]{0.05} & 
\makebox[0.3cm][r]{0.02} & \makebox[0.3cm][r]{1.9}  & 0.9 & 0.02 & 0.01 &  7\nl
\enddata
\label{tab-cocmed}
\end{deluxetable}

\begin{deluxetable}{lccccccccrr}
\tablecolumns{11}
\tablewidth{0pc}
\tablecaption{Average physical parameters}
\tablehead{
\colhead{\makebox[1cm][c]{Type}} & \colhead{M$_{r}$} & 
\colhead{L$_{H\alpha}$} & \colhead{z} &
\multicolumn{4}{c}{Equivalent width (\AA)} &
\colhead{E$_{\bv}$} & \colhead{$\bv$} & \colhead{$\vr$}  \nl 
\colhead{} & \colhead{} & \colhead{(10$^{\rm 8}$ L$_{\odot}$)} & \colhead{}
& \colhead{\makebox[1.0cm][c]{H$\alpha$ }} & 
\colhead{\makebox[1.0cm][c]{H$\beta$}} & \colhead{\makebox[1.0cm][c]{[OII]}} & 
\colhead{\makebox[1.0cm][c]{[OIII]}} & \colhead{} & \colhead{}
}
\startdata
Sy 1 & -21.3 & 5.22 & 0.0334 & 
\makebox[0.3cm][r]{260} & \makebox[0.3cm][r]{51} & 13 & 
\makebox[0.3cm][r]{38} & 0.463 & 0.10 & 0.26  \nl
Sy 2 & -22.0 & 3.78 & 0.0342 & 
\makebox[0.3cm][r]{86}  & \makebox[0.3cm][r]{9}  & 26 & 
\makebox[0.3cm][r]{90} & 0.794 & 0.47 & 0.56  \nl
SBN  & -21.1 & 2.24 & 0.0281 & 
\makebox[0.3cm][r]{77}  & \makebox[0.3cm][r]{9}  & 27 & 
\makebox[0.3cm][r]{11} & 0.789 & 0.30 & 0.30 \nl
DANS & -20.3 & 0.58 & 0.0243 & 
\makebox[0.3cm][r]{54}  & \makebox[0.3cm][r]{9}  & 38 & 
\makebox[0.3cm][r]{13} & 0.547 & 0.38 & 0.28  \nl
HIIH & -20.4 & 2.32 & 0.0240 & 
\makebox[0.3cm][r]{150} & \makebox[0.3cm][r]{24} &63 & 
\makebox[0.3cm][r]{74} & 0.514 & 0.04 & 0.06  \nl
DHIIH& -19.1 & 0.38 & 0.0226 & 
\makebox[0.3cm][r]{107} & \makebox[0.3cm][r]{17} & 64 & 
\makebox[0.3cm][r]{78} & 0.360 & 0.03 & 0.01 \nl
BCD  & -18.1 & 0.35 & 0.0226 & 
\makebox[0.3cm][r]{294} & \makebox[0.3cm][r]{72} & 89 & 
\makebox[0.3cm][r]{458}& 0.096 & -0.07&-0.09 \nl
\enddata
\label{tab-parmed}
\end{deluxetable}

\newpage

\begin{deluxetable}{lrrrrrr}
\tablewidth{0pc}
\tablecaption{[OIII] EW distributions}
\tablehead{
 & \multicolumn{2}{c}{UCM} & \multicolumn{2}{c}{KUG} & \multicolumn{2}{c}{UM}
}
\startdata
Total       & 145  &       & 105  &       & 139  &      \nl
No [OIII]   & 64   &(44\%) & 36   &(34\%) & 6    &(4\%) \nl
$<$ 10 \AA  & 47   &(32\%) & 35   &(33\%) & 21   &(15\%) \nl
$>$ 50 \AA  & 42   &(29\%) & 23   &(22\%) & 81   &(58\%) \nl
$>$ 100 \AA & 26   &(19\%) & 5    &(5\%)  & 61   &(44\%) \nl
$>$ 400 \AA & 5    &(3\%)  & 0    &(0\%)  & 23   &(16\%) \nl
\enddata
\label{tab-ewoiii}
\end{deluxetable}

\begin{deluxetable}{lrrrrrr}
\tablewidth{0pc}
\tablecaption{H$\beta$ EW distributions}
\tablehead{
 & \multicolumn{2}{c}{UCM} & \multicolumn{2}{c}{KUG} & \multicolumn{2}{c}{UM}
}
\startdata
Total        & 218 &       & 117    &      & 139 &       \nl
No H$\beta$  & 35  &(16\%) & \ldots &\ldots& 6   &(4\%)  \nl
$<$ 10 \AA   & 120 &(55\%) & 57     &(49\%)& 31  &(22\%) \nl
$>$ 40 \AA   & 22  &(10\%) & 2      &(2\%) & 41  &(29\%) \nl
$>$ 50 \AA   & 15  &(7\%)  & 1      &(1\%) & 33  &(24\%) \nl
\enddata
\label{tab-ewhb}
\end{deluxetable}

\begin{deluxetable}{lrrrrrrrrc}
\tablewidth{0pc}
\tablecaption{Distribution of the different ELG classes according to
metallicities and their mean values}
\tablehead{
\colhead{\makebox[2cm][c]{\small Type}} &\colhead{B} &
\colhead{C} &\colhead{D} &\colhead{E} & \colhead{F} & \colhead{{\small No}} & 
\colhead{$Z/Z_{\odot}$} & \colhead{$Z_{\odot}/Z$} & \colhead{log([O/H])+12} \nl
 &\colhead{Z$_{\odot}$} & \colhead{Z$_{\odot}$/2} &
\colhead{Z$_{\odot}$/5} &\colhead{Z$_{\odot}$/10} & 
\colhead{Z$_{\odot}$/20} & \colhead{{\small total}} & 
\colhead{{\small mean}} & \colhead{{\small mean}} & \colhead{{\small 
mean}} \nl
}
\startdata
SBN  &23 &35&3 &0 & 0 & 61  & 0.67  & 1.5 & 8.65 \nl
DANS &6  &11&0 &2 & 0 & 19  & 0.62  & 1.6 & 8.61 \nl
HIIH &5 & 8 & 19&3 &0 & 35  & 0.37  & 2.7 & 8.39 \nl
DHIIH&0 & 0 & 2 &11&1 & 14  & 0.11  & 9.0 & 7.87 \nl
BCD  &0 & 0 & 0&3  &3 &  6  & 0.075 & 13.0& 7.71 \nl
\enddata
\label{tab-abund}
\end{deluxetable}

\begin{deluxetable}{lrrccccc}
\tablewidth{0pc}
\tablecaption{Accurate abundances for selected UCM galaxies}
\tablehead{
\colhead{\makebox[2.5cm][c]{UCM}} & \colhead{N$_{e}$} &
\colhead{T$_{e}$} & \colhead{{\scriptsize 12+log(O/H)}} & \colhead{O/N} & 
\colhead{O$^{++}$/Ne$^{++}$} & \colhead{S$^{+}$/H$^{+}$} & \colhead{Y} \nl
 & \colhead{(cm$^{-3}$)} & \colhead{(10$^3$ K)} & & & & & \nl
}
\startdata
UCM0049$-$0006 & 100  & 15.5 & 7.76 & 0.60 & 0.79 & 5.36 & 0.265 \nl
UCM0049$+$0017 & 100  & 16.1 & 7.59 & 0.60 & 0.79 & 5.83 & 0.271 \nl
UCM0050$+$0005 & 30   & 13.1 & 8.10 & 1.25 & 0.58 & 5.86 & 0.254 \nl
UCM0056$+$0044 & 75   & 14.7 & 7.79 & 1.16 & 0.77 & 5.70 & 0.310 \nl
UCM0150$+$2032 & 200  & 10.0 & 8.22 & 1.08 & 0.63 & 5.90 & 0.248 \nl
UCM0156$+$2410 & 100  & 10.0 & 7.97 & 0.66 & 0.30 & 5.80 &       \nl
UCM1324$+$2926 & 120  & 14.5 & 7.86 & 1.21 & 0.70 & 5.70 & 0.295 \nl
UCM1331$+$2901 & 100  & 15.2 & 7.88 & 1.21 & 0.78 & 5.07 & 0.252 \nl
UCM1429$+$2645 & 30   & 16.5 & 7.80 & 0.83 & 0.75 & 5.83 &       \nl
UCM1612$+$1309 & 110  & 14.3 & 7.99 & 0.04 & 0.70 & 5.60 & 0.333 \nl
UCM2251$+$2405 & 200  & 9.9  & 7.94 & 0.55 & 0.77 & 6.10 &       \nl
UCM2304$+$1640 & 100  & 13.9 & 8.01 & 1.66 & 0.64 & 5.68 & 0.244 \nl
UCM2316$+$2028 & 100  & 10.0 & 8.01 &      & 0.85 &      &       \nl
UCM2326$+$2435 & 10   & 13.1 & 8.03 & 1.33 & 0.74 & 5.65 & 0.308 \nl
UCM2327$+$2515 & 180  & 14.7 & 7.90 & 1.21 & 0.49 & 5.77 &       \nl
\
\enddata
\label{tab-candida}
\end{deluxetable}

\begin{deluxetable}{ccccccccccc}
\tablewidth{0pc}
\tablecaption{Luminosity functions for all the four ELGs surveys}
\tablehead{
        & \multicolumn{2}{c}{UCM ELGs} &  \multicolumn{2}{c}{BST ELGs} &  
\multicolumn{2}{c}{UM ELGs} &
          \multicolumn{2}{c}{KUG ELGs} & \multicolumn{2}{c}{Case} \nl
\colhead{M$_{B}$} & \colhead{log$\phi$(M$_{B}$)} & \colhead{N}     
  & \colhead{log$\phi$(M$_{B}$)} & \colhead{N} &
\colhead{log$\phi$(M$_{B}$)} & \colhead{N}     &
\colhead{log$\phi$(M$_{B}$)} & \colhead{N}     &
\colhead{log$\phi$(M$_{B}$)}
& \colhead{N} \nl
}
\startdata
% MB     UCM   NUCM    BST  NBST   UM   NUM   KUG   NKUG    CASE   NCASE
 -23.0 & 0.00   & 0  & 0.00 & 0 & 0.00 & 0 & 0.00 & 0   & 0.00   & 0   \nl 
 -22.0 & -4.46  & 8  & 0.00 & 0 &-6.03 & 1 & -5.66 & 5   & -5.59  & 1   \nl
 -21.0 & -3.13  & 32 & 0.00 & 0 &-4.60 & 8  & -4.58 & 15  & -4.30  & 5  \nl
 -20.0 & -2.78  & 51 & 0.00 & 0 &-3.85 & 17 & -3.69 & 29  & -3.28  & 21 \nl
 -19.0 & -2.80  & 40 & 0.00 & 0 &-2.77 & 23 & -3.08 & 30  & -2.57  & 27 \nl
 -18.0 & -2.74  & 14 & -2.79& 1 &-2.94 & 24 & -3.12 & 7   & -2.37  & 10  \nl
 -17.0 & -2.43  & 7  & 0.00 & 0 &-2.88 & 15 & -2.74 & 4   & -1.92  & 9  \nl
 -16.0 & -3.43  & 1  & 0.00 & 0 &-1.76 & 21 & -2.06 & 5   & -1.79  & 4   \nl
 -15.0 & -3.29  & 1  & -1.57& 5 &-2.15 & 7  & -1.65 & 3   & -1.15  & 2   \nl
 -14.0 & 0.00   & 0  & -1.72& 2 &-2.67 & 4  & 0.00  & 0   & -1.06  & 1   \nl
 -13.0 & 0.00   & 0  & -1.82& 1 &-2.43 & 2  & 0.00  & 0   & 0.00   & 0   \nl
\enddata
\label{tab-lf}
\end{deluxetable}

\begin{deluxetable}{cccccc}
\tablewidth{0pc}
\tablecaption{Schechter functions for all the four ELGs surveys}
\tablehead{
 &\colhead{UCM}&\colhead{BST}&\colhead{UM}&\colhead{KUG}&\colhead{Case}\nl
}
\startdata
$M^{*}$       & -20.4  & \ldots    & -19.45   &  \ldots  &     -20.05  \nl
$\alpha$      & -0.90  & $\sim$-1.3&  -1.20   &  \ldots  &     -1.21   \nl
$\phi^{*}$    & 0.0033 & $<$0.007  &  0.0012  &  \ldots  &     0.0022  \nl
\enddata
\label{tab-fsch}
\end{deluxetable}

\clearpage

\figcaption[f1_diag.ps]{Line diagnostic diagram plotting [OIII]
$\lambda$5007/H$\beta$ against [NII] $\lambda$6584/H$\alpha$ both in
logarithmic scale. The different symbols indicate the ELG class (see
text).\label{f1-diag}}

\figcaption[f2_o3mb.ps]{Excitation versus absolute magnitude for the
UCM galaxies. The different symbols indicate the ELG class. They have
been also plotted the H$\alpha$ selected ELGs found by Boroson \etal (1993).
\label{f2-o3mb}}

\figcaption[f3_exci.ps]{Excitation versus [OII]/[OIII] ratio. The
symbols are the same as for Figure \ref{f1-diag}.\label{f3-exci}}

\figcaption[f4_ewoiii.ps]{Histogram of the [OIII] $\lambda$5007
EW in logarithmic
scale for the UCM, KUG and UM galaxies.\label{f4-ewoiii}}

\figcaption[f5_ewsurveys.ps]{Mean values, first quartiles and extreme
values for the [OIII] EWs at Case, UM, Kiso and UCM samples.
\label{f5-ewsurveys}}

\figcaption[f6_ewha.ps]{Histogram of the H$\alpha$+[NII] EW in logarithmic
scale for the UCM galaxies.\label{f6-ewha}}

\figcaption[f7_tartas.ps]{Spatial distribution for CfA (dots) and UCM
(open circles) galaxies.\label{f7-tartas}}

\figcaption[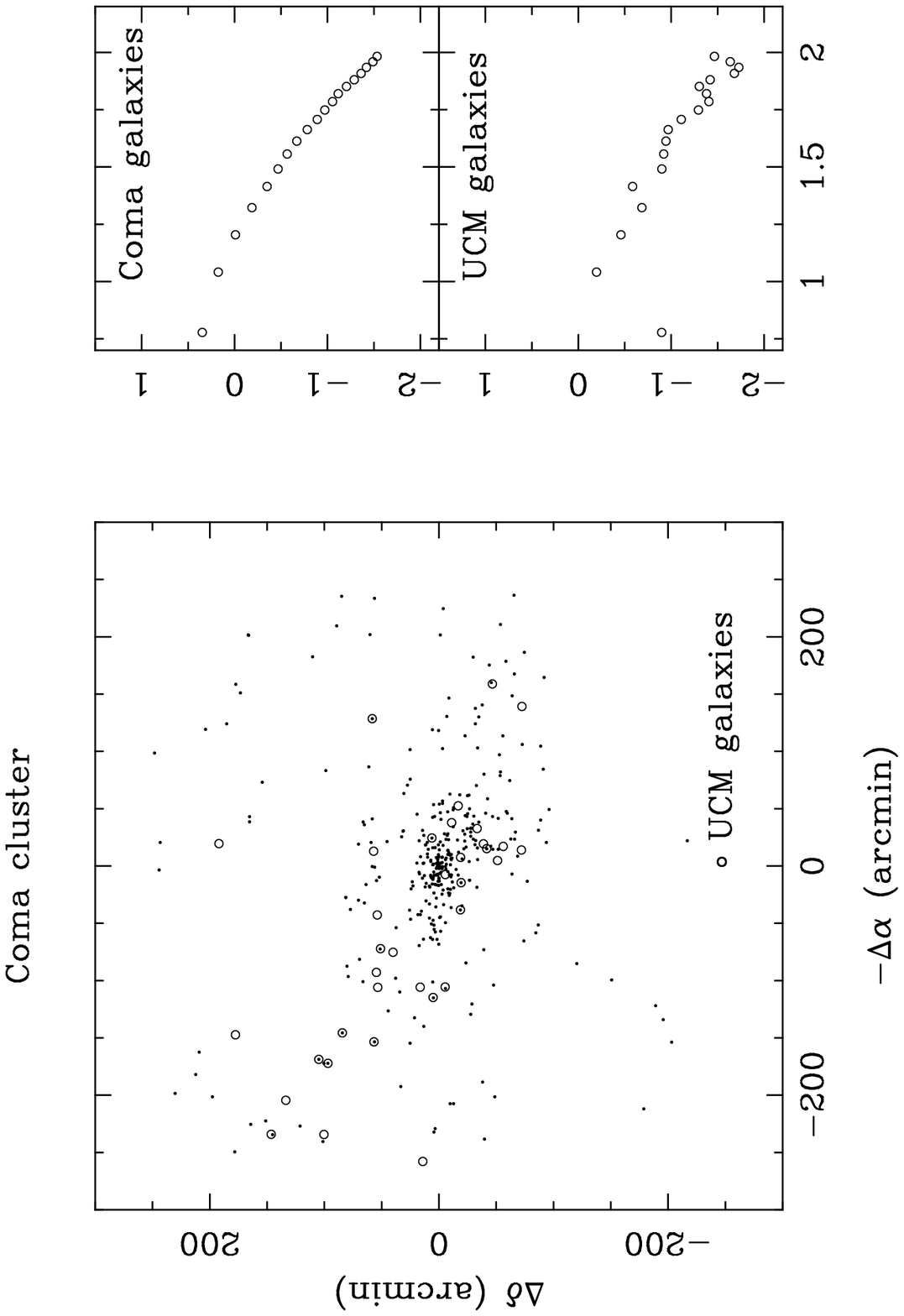]{Angular correlation function for Coma
galaxies and UCM galaxies in the cluster.\label{f8-cfa228}}

\figcaption[f9_ewhar.ps]{Logarithm of the H$\alpha$+[NII] equivalent
width versus r apparent magnitude. The lack of faint galaxies with low
EWs is clear (see text).\label{f9-ewhar}}

\figcaption[f10_ewhaz.ps]{Logarithm of the H$\alpha$+[NII] equivalent
width versus redshift. An asp correspond to the object a8.196. Other
objects with small emission but not detected are plotted as crosses.
\label{f10-ewhaz}}

\figcaption[f11_lf.ps]{Luminosity functions for the different ELGs surveys.
\label{f11-lf}}
\end{document}